# Title: Resonant-enhanced tunneling electroresistance in sliding ferroelectric tunnel junctions

**Authors:** Ruixue Wang[1,2,3,†], Jiangang Chen[1,†], Er Pan[1], Wunan Wang[2], Zefen Li[1], Fan Yang[1], Hongmiao Zhou[1,2], Zhaoren Xie[2], Qing Liu[1], Xiao Luo[1], Junhao Chu[2,3], Wenwu Li[2,*], and Fucai Liu[1,4,*]

[1]School of Optoelectronic Science and Engineering, University of Electronic Science and Technology of China, Chengdu 611731, China.

[2]Shanghai Frontiers Science Research Base of Intelligent Optoelectronics and Perception, Institute of Optoelectronics, College of Future Information Technology, Fudan University, Shanghai 200433, China.

[3]School of Physics, East China Normal University, Shanghai 200241, China.

[4]State Key Laboratory of Electronic Thin Films and Integrated Devices, University of Electronic Science and Technology of China, Chengdu 611731, China.

*Corresponding author. Email: fucailiu@uestc.edu.cn (F. L.); liwenwu@fudan.edu.cn (W. L.)

**Abstract:** The escalating demand for memory scaling requires switching mechanisms that remain reliable at atomic thickness while operating with minimal energy consumption. Sliding ferroelectricity provides a promising platform for this challenge: the spontaneous interfacial polarization emerging at superlubric, atomically thin van der Waals interfaces endows exceptional fatigue resistance, ultrafast switching and ultralow coercive fields. Nevertheless, the intrinsically weak polarization of sliding ferroelectrics limits the available signal window, necessitating new physical mechanisms that can transduce subtle polarization variations into pronounced resistance contrasts. Here, we address this challenge by introducing momentum-conserving resonant tunneling between lattice-aligned graphene electrodes. The resulting resonant sliding ferroelectric tunnel junction achieves a tunneling electroresistance (TER) ratio of up to 225.65%, substantially exceeding that of conventional sliding ferroelectric tunnel junctions. In addition, the device delivers a tunable TER ratio, multistate programmability, high current density, robust endurance with a small coefficient of variation (<0.69%), fast switching (20 ns), low switching energy (310 fJ), and low read voltage (<0.2 V). Collectively, these results establish a unique role for sliding ferroelectricity in bridging the gap of memory technology between performance and miniaturization, and open a new pathway toward next-generation nonvolatile memory technologies.

**Main Text:**

The push for aggressive memory scaling under data-intensive and artificial intelligence workloads confronts unavoidable reliability challenges (*1-4*). Existing memories encode information in degrees of freedom collinear with the writing field, rendering the thermal stability of the memory directly dependent on device dimensions along the field direction and placing information storage in fundamental conflict with scaling (*5-15*), as shown in Fig. 1A. Sliding ferroelectricity circumvents this dilemma by encoding states in discrete interlayer stacking configurations governed by in-plane atomic sliding orthogonal to the vertical writing field (*16-19*). This geometric orthogonality intrinsically decouples the information-carrying energy barrier from the scaling dimension (Fig. 1A), allowing sustained aggressive scaling without compromise. Meanwhile, the tunnel junction architecture allows further scaling by transforming quantum tunneling from a leakage path into a sensitive, non-destructive readout mechanism (*20-22*). While the atomic-scale thickness and superlubric interfaces (*23, 24*), combined with the outstanding fatigue resistance (*25, 26*), ultrafast programming speeds (*25, 27, 28*), and ultralow coercive fields of sliding ferroelectricity (Fig. 1B) (*23*), further eliminate critical barriers to extreme scaling without performance trade-offs. Thus, sliding-ferroelectric-based ferroelectric tunnel junctions (FTJs) emerge as a promising candidate for next-generation memory technology, uniquely positioned to bridge the gap between performance and miniaturization. However, the intrinsically weaker polarization of sliding ferroelectricity constrains the achievable tunneling electroresistance (TER) ratio (*29-32*), while a high TER ratio is essential for reliable state discrimination and noise-immune readout in scaled devices (Fig. 1B) (*33, 34*). Enhancing TER is thus critical to realize the full potential of sliding-ferroelectric FTJs for next-generation memory scaling.

Graphene resonant tunneling (*35, 36*) presents a promising solution to address this challenge. Since the density of states near the Dirac point is exceptionally low (*36-38*), neither graphene electrode can accumulate sufficient charge to compensate for the polarization-induced potential difference, resulting in a large shift in the relative band alignment between them upon ferroelectric polarization switching (*29, 39, 40*). This means that even the internal electric field generated by weak ferroelectric polarization effectively modulates the electrostatic potential across the junction. Meanwhile, such a change in the electrostatic potential modifies the energy alignment required for resonant tunneling, leading to a nonlinear shift of the resonance peak positions and switching the device between "resonant" and "off-resonant" states under specific bias conditions (Fig. 1C). This transition dramatically alters the tunneling probability through the junction, as transport in the resonant state is strongly enhanced due to momentum-conserving tunneling, whereas off-resonant transport exhibits greatly reduced transmission probability. Such a dramatic difference in tunneling probability, driven by quantum coherent transport, holds the potential to achieve an ultra-high TER response even within weakly polarized ferroelectric systems, thereby overcoming through the readout contrast bottleneck inherent in sliding ferroelectricity.

In this work, we designed and fabricated a resonant ferroelectric tunnel junction (R-FTJ) based on the single-layer graphene electrodes and a twisted bilayer boron nitride barrier. At room temperature, the TER (defined as $(R_{\text{off}}-R_{\text{on}})/R_{\text{on}}$ (*41, 42*)) reaches up to 225.65%, and can be flexibly tuned via polarization and gate voltage $V_{\text{g}}$. Furthermore, the device possesses multistate programmability, high current density, a switching speed of 20 ns, read voltages below 0.2 V, and a low switching energy of 310 fJ. It also exhibits excellent retention exceeding 10 years. Statistical analysis over 1000 switching cycles reveals that the coefficients of variation for the On and Off states are as low as 0.47% and 0.69%, respectively. These results highlight the strong potential of the proposed device in next-generation non-volatile memory architectures, delivering high reliability, low power consumption, and simultaneous compatibility with aggressive dimensional scaling and high-performance requirements.

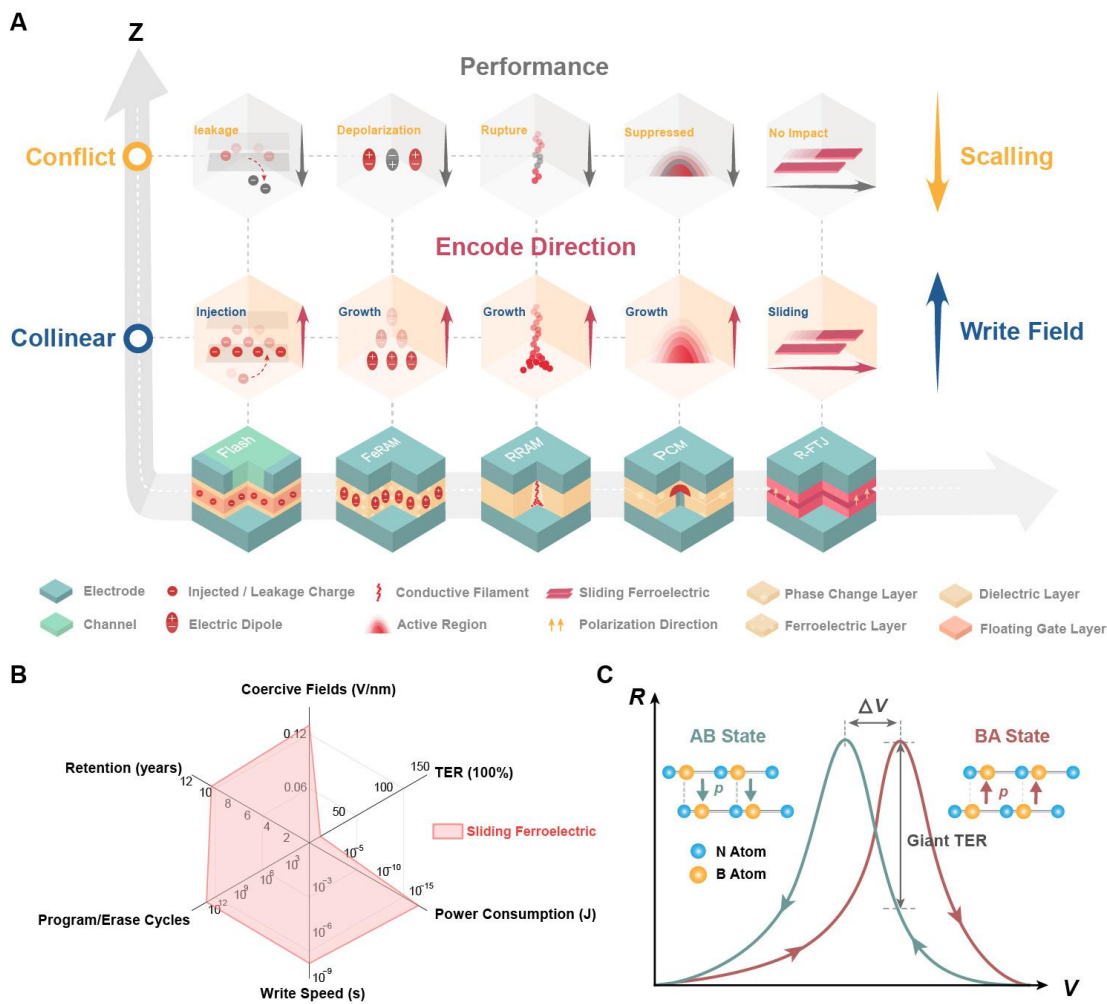

**Fig. 1. Schematic illustration of sliding ferroelectricity enabling extreme scaling without performance trade-offs and enhanced TER in R-FTJ**. **(A)** Comparison of encoding mechanisms and scaling performance in the Z direction across mainstream memory technologies and R-FTJ. Orange, blue and dark-grey arrows denote the write-field direction, encoding direction, and performance evolution under scaling, respectively. **(B)** Radar chart benchmarking key performance metrics of sliding ferroelectrics used in R-FTJ. **(C)** Schematic of the enhanced TER effect in R-FTJ enabled by graphene resonant tunneling. Here, ΔV represents the voltage shift due to polarization switching. Green and red arrows denote the polarization directions for the AB (left) and BA (right) stacking configurations of the sliding ferroelectricity, respectively.

**Resonant tunneling mechanism and enhanced TER**

The single-layer graphene/twisted-BN/single-layer graphene (slg/t-BN/slg) R-FTJ tunneling device we proposed is illustrated in Fig. 2A. Here, the twisted-BN as tunneling barrier is sandwiched between dual single-layer graphene (slg) electrodes, with the entire stack encapsulated by a thick top h-BN flake to preserve structural integrity and suppress environmental doping. In this configuration, the broken inversion symmetry induced by the moiré twist generates a robust out-of-plane spontaneous polarization ($P$) that correlates with the local atomic registry of AB and BA stacking configurations (Kelvin Probe Force Microscopy and Piezoresponse Force Microscopy characterization shown in Fig. S1). Notably, unlike conventional ferroelectric tunnel junctions that modulate resistance solely through barrier-height tuning, our device is designed to exploit resonant tunneling, enabling ferroelectric polarization to directly modulate quantum transport and thereby determine the resistance state.

The occurrence of resonant tunneling inherently requires simultaneous conservation of energy and momentum. In our device, the twist angle between the top and bottom graphene layers determines whether momentum conservation is satisfied (Fig. 2B), while energy conservation is controlled by tuning $V_g$ and $V_d$, as shown in Figs. 2C-E. In the momentum-matched state, the maximal overlap of the two Dirac cones enables the highest tunneling current (Fig. 2D), whereas misalignment of the Dirac cones reduces the current (Fig. 2E). On the other hand, when the top and bottom graphene lattices are misaligned, momentum mismatch suppresses resonant tunneling. Consequently, transport characteristics in the non-resonant sliding ferroelectric tunnel junction (slg/t-BN/slg NR-FTJ) device are primarily governed by ferroelectric polarization-induced barrier height modulation, exhibiting a smooth, non-linear $I_d$-$V_d$ curve, as shown in Fig. 2F. Even with the quantum capacitance effect (*36*), relying solely on barrier modulation renders the weak ferroelectric polarization insufficient to produce an appreciable hysteresis window between forward and backward scans, resulting in a maximum TER of only 32.63% at $V_g$=−40 V, with very limited tunability by gate voltage, as shown in Fig. S2.

In contrast, the slg/t-BN/slg R-FTJ introduces a new degree of freedom by enabling lattice alignment between the top and bottom graphene layers, transforming electron transport from conventional barrier-height modulation into a process in which momentum-conserving resonant tunneling renders conductance highly sensitive to interlayer band alignment. As shown in Fig. 2G, large hysteresis window and discrete current steps caused by polarization switching can be observed in the $I_d$-$V_d$ curve over the 0.2 V to 0.5 V range. The calculated TER value is 225.65%, which represents a substantial improvement (approximately two orders of magnitude higher) over the performance reported in existing sliding-ferroelectric-based tunneling junctions (Fig. 2H) (*43-45*).

In addition, in traditional ferroelectric tunnel junctions, the TER is typically limited by a fixed barrier height. However, in our resonant ferroelectric tunneling junction, the TER can be independently tuned by controlling the carrier concentration in graphene via $V_g$ or by modulating the polarization state. As shown in Fig. 2I, upon application of a positive $V_d$ pulse, the downward polarization reversal causes the $I_d$-$V_d$ curve to shift rightward, accompanied by an enlargement of the hysteresis window. Conversely, upon application of a negative $V_d$ pulse (upward polarization reversal), the $I_d$-$V_d$ curves shifts leftward, and the hysteresis window narrows accordingly. The gate voltage dependence of the $I_d$-$V_d$ curves is shown in Fig. 2J. As $V_g$ is swept from -50 V to 50 V, the resonance peak current exhibits a clear gate-dependent modulation, accompanied by a pronounced shift of the resonance peak along the voltage axis. The resulting TER as a function of the polarization state and $V_g$ is summarized in Fig. 2K, where it reaches a maximum at $V_g$=0 V and a $V_d$ pulse amplitude of -2 V.

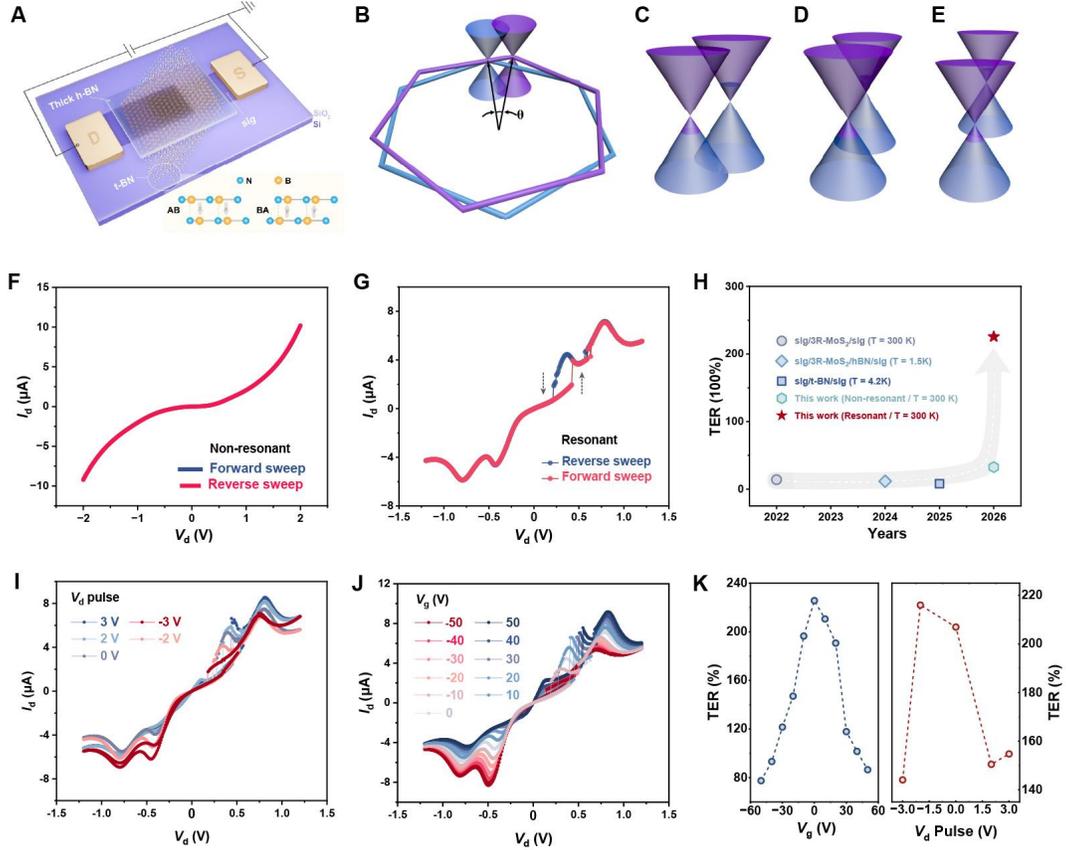

**Fig. 2. Structural and ferroelectric characterization of the slg/t-BN/slg R-FTJ.** (**A**) Schematic illustration of the device structure. The stack consists of twisted hBN sandwiched between two graphene electrodes, encapsulated by a top h-BN flake and supported by a Si/SiO$_2$ substrate. The inset represents the AB and BA stacking configurations. The red and blue spheres denote Nitrogen (N) and Boron (B) atoms, respectively. The gray arrows marked with P indicate the direction of the out-of-plane electric polarization induced by the specific stacking symmetry. (**B**) Schematic representation of the twist angle θ in momentum space between the top and bottom graphene layers. (**C**)-(**E**) Schematic diagrams showing the evolution of Dirac cones under various energy level alignment conditions, corresponding to the resonant tunneling process. (**F**) The tunneling $I_d$-$V_d$ characteristics of the slg/t-BN/slg NR-FTJ at $V_g$=0 V. (**G**) The tunneling $I_d$-$V_d$ characteristics of the slg/t-BN/slg R-FTJ at $V_g$=0 V. (**H**) Comparison of TER values between this work and similar devices in recent years (*43-45*). (**I**) The $I_d$-$V_d$ characteristic curves under different $V_d$ pulses of different amplitudes with a pulse width of 0.01 s. (**J**) The $I_d$-$V_d$ characteristic curves under different gate voltages. $V_g$ varies from -50 V to 50 V (with a step of 10 V). (**K**) The TER values at different $V_d$ pulse and $V_g$ were calculated from the (I) and (J), respectively.

To evaluate the robustness and stability of the enhanced TER, consecutive voltage sweeps were performed at $V_g$=0 V and $V_g$=-10 V (Fig. S3). Despite repeated scanning cycles, the current curves remain highly consistent, exhibiting nearly identical traces across all measurements. Neither the positions of the current jumps nor the resonance peaks show any noticeable variation. This high reproducibility indicates the structural robustness of the device under repeated electric-field cycling and provides a reliable basis for quantitative evaluation of polarization-induced resistance modulation. In addition, measurements of the $I_d$-$V_d$ characteristics at different sweep rates (Fig. S4) rule out non-ferroelectric hysteresis arising from charge trapping. These results indicate that the observed phenomena are attributed to intrinsic behaviors dominated by ferroelectric polarization.

**Origin of the enhanced TER**

The enhanced TER in our device arises from the coupling between resonant tunneling and sliding ferroelectric polarization dynamics. As shown in Fig. 3A, polarization switching in the sliding ferroelectric layer occurs via dynamic evolution of AB and BA stacking domains. At low bias ($V_d \approx 0$ V), the system remains in a multi-domain configuration. As the bias increases, individual ferroelectric domains progressively overcome the sliding energy barriers associated with domain wall pinning, leading to transitions between AB and BA stacking states. This discrete domain evolution manifests electrically as step-like current jumps in the $I_d$-$V_d$ characteristics (measured by applying a triangular-pulse electric field waveform, Fig. S5A), as shown in Fig. 3B. Because AB and BA domains possess opposite polarization orientations, the built-in electric fields shift the relative positions of the Dirac cones in the two graphene electrodes. This shift alters the energy alignment, inducing transitions between resonant and off-resonant tunneling regimes (Fig. 3C). In this process, the bias required to align the Dirac cones varies substantially, leading to distinct shifts in the resonance peak positions along the voltage axis. As shown in Fig. 3D, the resonance peaks appear at different positions during forward and backward sweeps, giving rise to pronounced resistance hysteresis associated with polarization history. Under a given bias condition, TER is therefore governed by differences in the resonant conditions determined by the state of polarization. When opposite polarization directions correspond to resonant and off-resonant tunneling regimes, the tunneling probability differs substantially, generating a large resistance contrast and thus a strongly enhanced TER.

In addition, $V_d$ pulses with different amplitudes can program the ferroelectric domain configuration, thereby defining the polarization state and the corresponding resonant condition of the device. As a result, different polarization states lead to distinct tunneling responses under a given read bias, enabling effective control of TER. Meanwhile, the gate voltage $V_g$ tunes the carrier concentration and Fermi level in graphene, further adjusting the band alignment between the electrodes. For a given polarization state, different $V_g$ values tune different resonant conditions, altering the current contrast between resonant and off-resonant states and thereby modulating TER. Finally, TER is jointly determined by the polarization state written by the $V_d$ pulse and the $V_g$-controlled band alignment, providing additional degrees of freedom to optimize the readout window, operating voltage, and energy consumption.

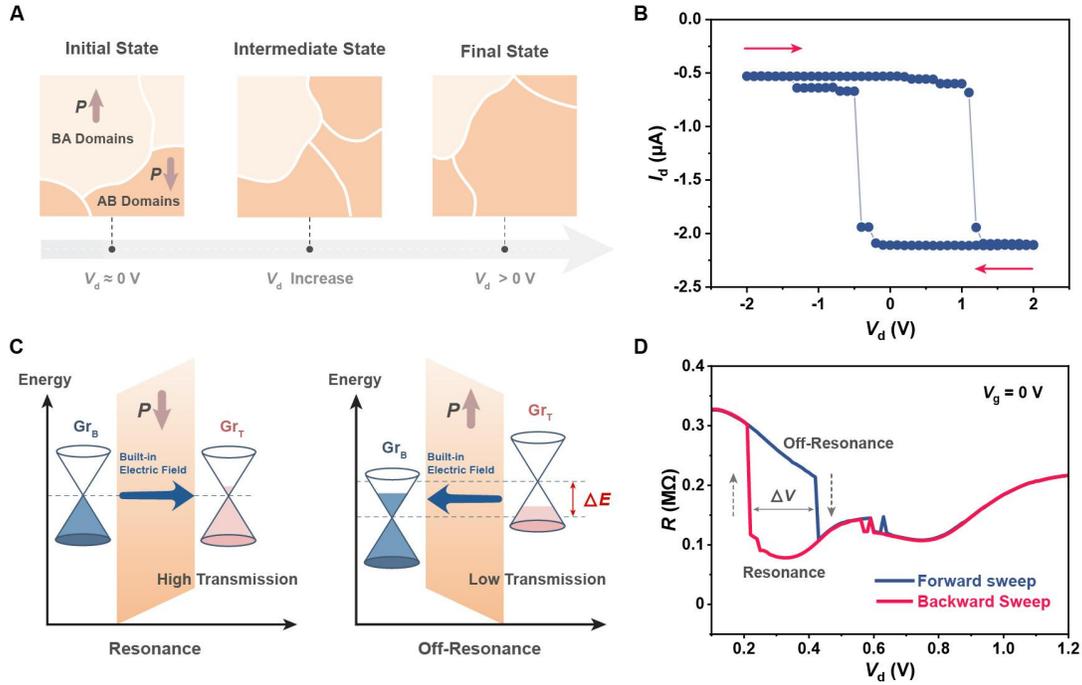

**Fig. 3. Schematic illustration of resonant tunneling-enhanced TER and ferroelectric domain dynamics. (A)** Schematic illustration of the evolution of AB and BA stacking domains in the sliding ferroelectric layer. At zero bias ($V_d$), multiple domains coexist with opposite polarization directions ($P$). As $V_d$ increases, domains progressively overcome the sliding energy barrier, evolving through intermediate states to a final configuration dominated by a single domain type. **(B)** The typical ferroelectric electrical transport measurements, where the $I_d$ as a function of applied voltage $V_d$. **(C)** Schematic energy diagrams of the graphene-ferroelectric-graphene junction. Left: polarization pointing down aligns the Dirac cones, enabling resonant tunnelling and high transmission. Right: polarization pointing up misaligns the Dirac cones, producing off-resonance conditions and low transmission. The built-in electric field generated by the ferroelectric polarization shifts the relative band positions. **(D)** The device resistance ($R$) versus $V_d$ at zero gate bias ($V_g$=0 V). Forward (blue) and backward (red) sweeps show pronounced hysteresis (ΔV) and peak shifts between resonance and off-resonance states.

**Non-volatile memory performance and reliability of the device**

To more comprehensively assess the device performance and to verify that the achievement of the enhanced TER does not come at the expense of other key metrics, we performed systematic electrical transport measurements. First, we measured the switched fraction as a function of pulse width for several excitation voltages, with 10 switching cycles recorded for each combination of pulse width and amplitude (Fig. 4A). Here, our pulse-generation and measurement setup is limited to a minimum pulse width of approximately 20 ns. This result indicates that the device we proposed allows nanosecond-regime write/erase operations, without any compromise in TER or reproducibility.

The non-volatile nature of the On and Off states was further confirmed through long-term retention measurements as shown in Fig. 4B. After the device was set into the respective logic states with a 1 ms pulse, the current levels were monitored at a non-disruptive read voltage ($V_{read}$=-0.1 V). The temporal evolution of the current exhibits negligible relaxation over $10^4$ s. By extrapolating the experimental retention curves, the device is projected to maintain its memory states for over 10 years, which meets the requirements for industrial non-volatile storage. Moreover, discrete intermediate resistance states can be achieved under different programming conditions (Fig. S6), demonstrating its potential of the device for multilevel data storage. Meanwhile, the switching behavior in static electrical measurements remains highly consistent even after 100 consecutive measurement cycles (Fig. S5B), underscoring the exceptional operational robustness and reproducibility of the device.

The device also exhibits exceptional endurance and operational stability. The robust cycling performance is statistically corroborated by the histogram analysis shown in Fig. 4C. The current distributions corresponding to programming state exhibit clear separation, with the calculated coefficients of variation ($C_v$) as low as 0.47% for the On-state and 0.69% for the Off-state, underscoring the high degree of programming accuracy and reproducibility of the device. Collectively, these results demonstrate that an enhanced TER can be realized in our device without compromising other essential device characteristics, enabling a reliable platform for high-density, low-power memory architectures.

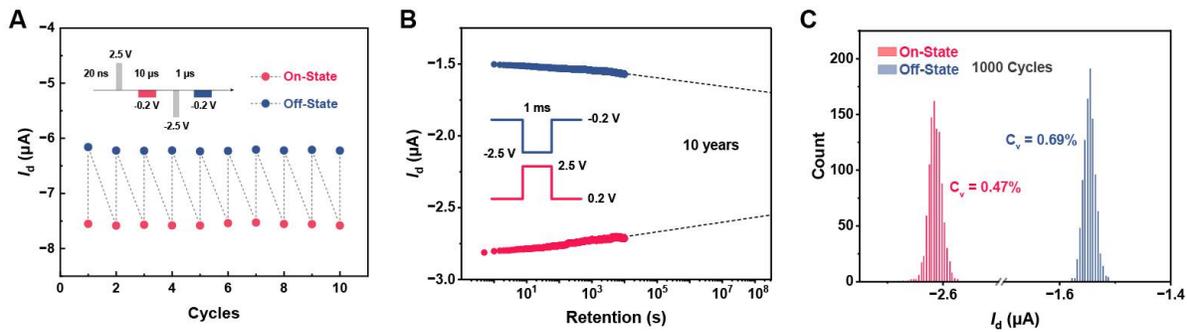

**Fig. 4. Non-volatile memory performance and reliability of the slg/t-BN/slg R-FTJ. (A)** Cyclic switching of the drain current $I_d$ between the On and Off states over ten operation cycles, demonstrating reproducible resistive switching induced by the applied voltage pulses with an amplitude of 2.5 V (-2.5 V) and a pulse width of 20 ns, as illustrated in the inset. **(B)** Data retention characteristics. After applying programming/erasing pulses of 2.5 V, the high and low current states remain stable for over $10^4$ s. Extrapolation using dashed lines indicates a retention potential exceeding 10 years at room temperature. **(C)** Statistical distribution of $I_d$ for the On and Off states after 1000 switching cycles. The two well-separated current distributions with small coefficients of variation ($C_v$ = 0.47% for On state and 0.69% for Off state) indicate reliable state discrimination and good device uniformity.

**Disscussion**

In summary, we demonstrate that sliding ferroelectricity and resonant tunneling form a natural combination for next-generation non-volatile memory. Sliding ferroelectricity provides an atomically thin interfacial polarization configuration with excellent fatigue resistance and ultralow switching power. Its atomic-scale thickness favors momentum-conserving transmission, well-defined stacking configurations create reproducible built-in electric fields that shift the resonance condition in a controlled way, and low coercive fields enable reversible switching with minimal electrical perturbation. Together, these attributes preserve the sharp energy selectivity that underlies the high-contrast readout enabled by resonant tunneling. Resonant tunneling, in turn, enables high-contrast electrical readout of weak sliding ferroelectric states because small polarization-induced potential variations drive the junction into and out of resonance, causing large changes in tunneling probability and resulting in pronounced resistance contrasts.

The resulting resonant sliding ferroelectric tunnel junction delivers a TER ratio of up to 225.65%, multistate programmability, high current density, robust endurance with a coefficient of variation as low as 0.69%, a 20 ns switching speed, sub-0.2 V read voltages, retention exceeding 10 years, and a low operating energy of 310 fJ. These results demonstrate that our proposed architecture not only retains the inherent advantages of sliding ferroelectricity, but also leverages resonant tunneling as a new physical control dimension to overcome the weak polarization readout challenge in sliding ferroelectricity. More importantly, this work points to a new direction for exploring the coupling of ferroelectricity with multi-physics effects at the ultimate thickness limit. Furthermore, given the layer-dependent polarization properties in multilayer sliding ferroelectricity (*45*), we foresee that the layer count offers an effective means to engineer the desired coupling strength between ferroelectric polarization and resonant tunneling, thereby providing a viable route toward customized device functionality. Meanwhile, super-lubricant van der Waals cavity arrays are expected to facilitate the practical applicability of our resonant tunneling devices in next-generation nanoelectronics (*46*).

**Acknowledgments**

This work was supported by the National Natural Science Foundation of China (92477115, U25A20482, 62374043), The Scientific Research Innovation Capability Support Project for Young Faculty (ZYGXQNJSKYCXNLZCXM-I7), Shanghai Pilot Program for Basic Research-Fudan University 21TQ1400100 (25TQ001), Shanghai Science and Technology Committee (Grant No. 25JD1400900), Sichuan Science and Technology Program (2025ZYD0182), Sichuan Province Key Laboratory of Display Science and Technology, and National Key Laboratory of Integrated Circuit Materials (SKLJC-K2025-04), State Key Laboratory of Dynamic Measurement Technology, North University of China (2024-SYSJJ-06).

**References and Notes**

1. K. Ishimaru, in *2019 IEEE International Electron Devices Meeting (IEDM)*. (2019), pp. 1.3.1-1.3.6. doi:10.1109/IEDM19573.2019.8993609

2. Q. Xia, J. J. Yang, Memristive crossbar arrays for brain-inspired computing. *Nat Mater* 18, 309-323 (2019). doi:10.1038/s41563-019-0291-x

3. A. Sebastian, M. Le Gallo, R. Khaddam-Aljameh, E. Eleftheriou, Memory devices and applications for in-memory computing. *Nat Nanotechnol* 15, 529-544 (2020). doi:10.1038/s41565-020-0655-z

4. A. Gholami *et al.*, AI and Memory Wall. *IEEE Micro* 44, 33-39 (2024). doi:10.1109/MM.2024.3373763

5. Y. Xiang *et al.*, Subnanosecond flash memory enabled by 2D-enhanced hot-carrier injection. *Nature* 641, 90-97 (2025). doi:1038/s41586-025-08839-w

6. M. Trieloff *et al.*, Structure and thermal history of the H-chondrite parent asteroid revealed by thermochronometry. *Nature* 422, 502-506 (2003). doi:10.1038/nature01499

7. Y. Jiang *et al.*, A scalable integration process for ultrafast two-dimensional flash memory. *Nature Electronics* 7, 868-875 (2024). doi:10.1038/s41928-024-01229-6

8. A. Makarov, V. Sverdlov, S. Selberherr, Emerging memory technologies: Trends, challenges, and modeling methods. *Microelectron Reliab* 52, 628-634 (2012). doi:10.1016/j.microrel.2011.10.020

9. I. P. Batra, B. D. Silverman, Thermodynamic stability of thin ferroelectric films. *Solid State Communications* 11, 291-294 (1972). doi:10.1016/0038-1098(72)91180-5

10. B. Shen *et al.*, Review on Ferroelectricity and Atomic Characterization of $Hf_{0.5}Zr_{0.5}O_2$ in FeRAM. *Acs Appl Electron Ma* 7, 4675-4702 (2025). doi:10.1021/acsaelm.5c00037

11. P. Noé, C. Vallée, F. Hippert, F. Fillot, J.-Y. Raty, Phase-change materials for non-volatile memory devices: from technological challenges to materials science issues. *Semicond Sci Tech* 33, 013002 (2017). doi:10.1088/1361-6641/aa7c25

12. T. Kim, S. Lee, Evolution of Phase-Change Memory for the Storage-Class Memory and Beyond. *Ieee T Electron Dev* 67, 1394-1406 (2020). doi:10.1109/ted.2020.2964640

13. S. K. Lai, Flash memories: Successes and challenges. *IBM Journal of Research and Development* 52, 529-535 (2008). doi:10.1147/rd.524.0529

14. H. Hwang, S. Youn, H. Kim, Recent advances in ferroelectric materials, devices, and in-memory computing applications. *Nano Converg* 12, 55 (2025). doi:10.1186/s40580-025-00520-2

15. A. Chen, A review of emerging non-volatile memory (NVM) technologies and applications. *Solid State Electron* 125, 25-38 (2016). doi:10.1016/j.sse.2016.07.006

16. A. Weston *et al.*, Interfacial ferroelectricity in marginally twisted 2D semiconductors. *Nature Nanotechnology* 17, 390-395 (2022). doi:10.1038/s41565-022-01072-w

17. K. Yasuda, X. Wang, K. Watanabe, T. Taniguchi, P. Jarillo-Herrero, Stacking-engineered ferroelectricity in bilayer boron nitride. *Science* 372, 1458-1462 (2021). doi: 10.1126/science.abd3230

18. M. Wu, J. Li, Sliding ferroelectricity in 2D van der Waals materials: Related physics and future opportunities. *Proceedings of the National Academy of Sciences* 118, e2115703118 (2021). doi: 10.1073/pnas.2115703118

19. F. Sui *et al.*, Atomic-level polarization reversal in sliding ferroelectric semiconductors. *Nat Commun* 15, 3799 (2024). doi:10.1038/s41467-024-48218-z

20. V. Garcia *et al.*, Giant tunnel electroresistance for non-destructive readout of ferroelectric states. *Nature* 460, 81-84 (2009). doi:10.1038/nature08128

21. Z. Xi *et al.*, Giant tunnelling electroresistance in metal/ferroelectric/semiconductor tunnel junctions by engineering the Schottky barrier. *Nat Commun* 8, 15217 (2017). doi:10.1038/ncomms15217

22. V. Garcia, M. Bibes, Ferroelectric tunnel junctions for information storage and processing. *Nat Commun* 5, 4289 (2014). doi:10.1038/ncomms5289

23. M. Vizner Stern *et al.*, Interfacial ferroelectricity by van der Waals sliding. *Science* 372, 1462-1466 (2021). doi:10.1126/science.abe8177

24. Z. Zheng *et al.*, Unconventional ferroelectricity in moire heterostructures. *Nature* 588, 71-76 (2020). doi:10.1038/s41586-020-2970-9

25. K. Yasuda *et al.*, Ultrafast high-endurance memory based on sliding ferroelectrics. *Science* 385, 53-56 (2024). doi:10.1126/science.adp3575

26. R. Bian *et al.*, Developing fatigue-resistant ferroelectrics using interlayer sliding switching. *Science* 385, 57-62 (2024). doi:10.1126/science.ado1744

27. Y. Bai *et al.*, Sub-nanosecond polarization switching with anomalous kinetics in vdW ferroelectric $WTe_2$. *Nat Commun* 16, 7221 (2025). doi:10.1038/s41467-025-62608-x

28. J. Liang *et al.*, Nanosecond Ferroelectric Switching of Intralayer Excitons in Bilayer 3R−$MoS_2$ through Coulomb Engineering. *Physical Review X* 15, 021081 (2025). doi:10.1103/PhysRevX.15.021081

29. X. Lv, T. Taniguchi, K. Watanabe, M. Lv, J. Xue, Electronic density of states modulated sliding ferroelectric tunneling junctions. *Phys Rev B* 112, 235434 (2025). doi:10.1103/bqcm-j3sh

30. M. Y. Zhuravlev, R. F. Sabirianov, S. S. Jaswal, E. Y. Tsymbal, Giant Electroresistance in Ferroelectric Tunnel Junctions. *Phys Rev Lett* 94, 246802 (2005). doi:10.1103/PhysRevLett.94.246802

31. Z. Wang, Z. Gui, L. Huang, Sliding ferroelectricity in bilayer honeycomb structures: A first-principles study. *Phys Rev B* 107, 035426 (2023). doi:10.1103/PhysRevB.107.035426

32. Y. Yan, M. Wu, Ionic Sliding Ferroelectricity in Layered Ion Conductors. *Phys Rev Lett* 135, 236801 (2025). doi:10.1103/fgxy-cgmh

33. J. Wu *et al.*, High tunnelling electroresistance in a ferroelectric van der Waals heterojunction via giant barrier height modulation. *Nature Electronics* 3, 466-472 (2020). doi:10.1038/s41928-020-0441-9

34. X. R. Wang, J. L. Wang, Ferroelectric tunnel junctions with high tunnelling electroresistance. *Nature Electronics* 3, 440-441 (2020). doi:10.1038/s41928-020-0463-3

35. Z. Zhang *et al.*, Toward High-Peak-to-Valley-Ratio Graphene Resonant Tunneling Diodes. *Nano Lett* 23, 8132-8139 (2023). doi:10.1021/acs.nanolett.3c02281

36. L. Britnell *et al.*, Resonant tunnelling and negative differential conductance in graphene transistors. *Nat Commun* 4, 1794 (2013). doi:10.1038/ncomms2817

37. J. Martin *et al.*, Observation of electron–hole puddles in graphene using a scanning single-electron transistor. *Nat Phys* 4, 144-148 (2007). doi:10.1038/nphys781

38. R. Moriya *et al.*, Influence of the density of states of graphene on the transport properties of graphene/$MoS_2$/metal vertical field-effect transistors. *Appl Phys Lett* 106, 223103 (2015). doi:10.1063/1.4921920

39. D. Koprivica, E. Sela, Resonant tunneling in graphene-ferroelectric-graphene junctions. *Phys Rev B* 106, 144110 (2022). doi:10.1103/PhysRevB.106.144110

40. Z. Fei *et al.*, Ferroelectric switching of a two-dimensional metal. *Nature* 560, 336-339 (2018). doi:10.1038/s41586-018-0336-3

41. Y. W. Yin *et al.*, Enhanced tunnelling electroresistance effect due to a ferroelectrically induced phase transition at a magnetic complex oxide interface. *Nat Mater* 12, 397-402 (2013). doi:10.1038/nmat3564

42. S. S. Parkin *et al.*, Giant tunnelling magnetoresistance at room temperature with MgO (100) tunnel barriers. *Nat Mater* 3, 862-867 (2004). doi:10.1038/nmat1256

43. Y. Gao *et al.*, Tunnel junctions based on interfacial two dimensional ferroelectrics. *Nat Commun* 15, 4449 (2024). doi:10.1038/s41467-024-48634-1

44. B. Vareskic *et al.*, Gate-Tunable Electroresistance in a Sliding Ferroelectric Tunnel Junction. *Nano Lett* 25, 17540-17546 (2025). doi:10.1021/acs.nanolett.5c03367

45. P. Meng *et al.*, Sliding induced multiple polarization states in two-dimensional ferroelectrics. *Nat Commun* 13, 7696 (2022). doi:10.1038/s41467-022-35339-6

46. Y. Yeo *et al.*, Polytype switching by super-lubricant van der Waals cavity arrays. *Nature* 638, 389-393 (2025). doi:10.1038/s41586-024-08380-2